\documentclass[prb,showpacs,twocolumn,superscriptaddress,amsmath,amssymb]{revtex4-1}
\date{today}

\usepackage{graphicx}
\usepackage[colorlinks=true,citecolor=blue]{hyperref}
\usepackage{hypernat}
\usepackage{units}
\usepackage{verbatim}

\begin{document}

\title{Driven superconducting proximity effect in interacting quantum dots}
\author{Ali G. Moghaddam}
\affiliation{Theoretische Physik, Universit\"at Duisburg-Essen and CeNIDE, 47048 Duisburg, Germany}
\author{Michele Governale}
\affiliation{School of Physical and Chemical Sciences and MacDiarmid
Institute for Advanced Materials and Nanotechnology, Victoria
University of Wellington, Wellington 6140, New
Zealand}
\author{J\"urgen K\"onig}
\affiliation{Theoretische Physik, Universit\"at Duisburg-Essen and CeNIDE, 47048 Duisburg, Germany}

\date{\today}

\begin{abstract}
We present a theory of non-equilibrium superconducting proximity effect in an interacting quantum dot induced by a time-dependent tunnel coupling between dot and a superconducting lead. 
The proximity effect, that is established when the driving frequency fulfills a gate-voltage-dependent resonance condition, can be probed through the tunneling current into a weakly-coupled normal lead.
Furthermore, we propose to generate and manipulate coherent superpositions of quantum-dot states with electron numbers differing by two, by applying pulsed oscillatory variations to the couplings between dot and superconductors. 
\end{abstract}

\maketitle

\section{Introduction}
The possibility to induce superconducting correlations in a quantum dot by 
the proximity of superconducting leads has received substantial interest in the last decade.\cite{de franceschi, 
yeyati11}
A variety of transport phenomena in hybrid normal/superconducting/quantum-dot systems have been addressed experimentally, including the Josephson effect,\cite{nanotube} $\pi$-junction behavior,\cite{kouwenhoven06, lindelof07} superconducting quantum interference,\cite{monthioux06} Cooper pair splitting,\cite{schonenberger09,strunk10,recher01} and Andreev-level spectroscopy.\cite{joyez10,deacon10} Theoretical studies have focused  on the interplay of superconductivity with the Kondo effect,\cite{ambegaokar00,avishai03,choi04,egger04,sellier05,lopez07} nonequilibrium Andreev transport,\cite{fazio98, kang98, schwab99, clerk00, cuevas01,futterer09,eldridge10,braggio11} and multiple Andreev reflection.\cite{yeyati97,schonenberger03,egger08}
\par
In equilibrium, strong Coulomb repulsion in a quantum dot suppresses the formation of a pair amplitude, while different-time superconducting correlations are induced by higher-order tunneling. 
As a consequence, the Josephson current through a quantum dot tunnel-coupled to two superconductors will be at least of second order in the tunneling rate $\Gamma$.\cite{glazman89,rozhkov01}
Experiments with quantum dots realized in carbon nanotubes\cite{nanotube,lindelof07,monthioux06} and InAs nanowires\cite{kouwenhoven06} are consistent with this scenario. However, it has been proposed that a non-equilibrium occupation of the quantum dot's states, induced by a voltage-biased normal lead, can lead to a finite pair amplitude on the dot.\cite{pala07,governale08}
In this Article, we propose an alternative way, namely using tunnel couplings that oscillate in time, to establish a non-equilibrium situation that supports superconducting correlations in the dot. 
\par
Transport through few-electron quantum dots can often be described by taking into account only a single orbital level.
In this case, a finite pair amplitude in the dot describes a coherent superposition of this level being empty, denoted by $|0\rangle$, and doubly occupied $|d\rangle$. 
It can be generated by driving the system via oscillatory tunnel couplings.
The resonance condition under which the pair amplitude becomes maximal involves the oscillation frequency and the energy difference $\delta$ between the  $|d\rangle$ and $|0\rangle$ states. 
The superconducting proximity effect can be probed by measuring the current into an additional, normal lead weakly coupled to the dot. Finally, we show that suitable pulsed oscillatory variations of the system parameters can produce any coherent superpositions of the $|0\rangle$ and $|d\rangle$ states. 
\par
\section{Model}
We consider a single-level quantum dot tunnel coupled to superconducting and normal leads, which can be described by the Anderson-impurity model with the Hamiltonian
$H=H_{\rm QD}+\sum_{\eta=S,N}(H_{\eta}+H_{T,\eta})$. 
We denote with $d^{\dagger}_{\sigma} (d_{\sigma})$ and
$c^{\dagger}_{\eta k \sigma}$ ($c_{\eta k \sigma}$) the  creation (annihilation) operators for electrons with spin $\sigma=\uparrow,\downarrow$ in the dot and for single-particle states with quantum number $k$ in lead $\eta$, respectively.
The quantum-dot Hamiltonian reads $H_{\rm QD} = \sum_{\sigma}\epsilon d^{\dagger}_{\sigma} d_{\sigma}+Ud^{\dagger}_{\uparrow} d^{\dagger}_{\downarrow}d_{\downarrow} d_{\uparrow}$, where $\epsilon$ and $U$ are the  single-particle energy and the  Coulomb energy for double occupation, respectively.
The leads are described by the BCS Hamiltonian $H_{\eta}=\sum_{k \sigma} 
\epsilon_{k}c_{\eta k \sigma}^\dagger c_{\eta k \sigma} - \sum_k ( 
\Delta_\eta c_{\eta k \uparrow}^\dagger c_{\eta -k \downarrow}^\dagger 
+{\rm H.c.} )$, with the superconducting pair potential $\Delta_{\eta}=|\Delta|e^{i\phi_{\eta}}$ being non zero only for the  superconductors.
The coupling between dot and lead $\eta$ is captured by the standard tunneling Hamiltonian $H_{T,\eta}= V_{\eta} \sum_{k \sigma} ( c_{\eta k \sigma}^\dagger 
d_\sigma +{\rm H.c.} )$. We assume the tunnel matrix elements $V_\eta$ to be independent of $k$ and $\sigma$, and define the tunnel-coupling strength as $\Gamma_{\eta}=2\pi |V_{\eta}|^2\sum_{k}\delta(\omega-\epsilon_k)$.  
\par
Since we focus on the proximity effect, whose microscopic origin is sub-gap Andreev reflection, we consider the limit of large superconducting gap, $|\Delta|\rightarrow \infty$. 
In this limit, the tunnel coupling to the superconducting leads can be taken into account by an effective dot Hamiltonian
$H_{\rm QD,eff}= H_{\rm QD} -{\Delta}_{\rm eff} (t)d^{\dag}_{\uparrow}d^{\dag}_{\downarrow}-{\Delta}_{\rm eff} ^{\ast} (t) d_{\downarrow}d_{\uparrow}$, that shows a BCS-like term ${\Delta}_{\rm eff}$,  whose strength is determined by the tunnel couplings to the superconducting leads. 
The effective-Hamiltonian formalism has been derived and employed in the time-independent case,\cite{governale08,rozhkov00} but it remains valid in the time-dependent case since non-instantaneous contributions play a role only for frequencies larger than the superconducting gap $|\Delta|$. 
\begin{figure}[tp]
	\includegraphics[width=7cm]{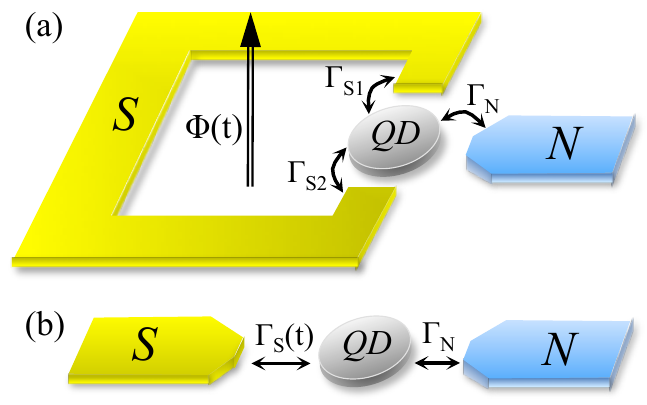}
	\caption{\label{fig1}(Color online) Schematic setups. 
	The oscillation in the effective coupling between dot and superconductor can be realized by (a) a 
	time-dependent phase difference between the superconductors or (b) varying the tunnel-coupling strength between dot and superconductor.}
\end{figure}
\par
We propose two different schemes to generate a time-dependent coupling ${\Delta}_{\rm eff}(t)$. In the first scheme, shown in Fig.~ \ref{fig1} (a), the dot is attached to one normal lead and two superconducting leads with a 
time-dependent phase difference $\Phi(t)=2\Omega t$.   
The latter can be created by monotonically increasing the magnetic flux enclosed by the superconducting loop or by directly applying a voltage $V=\hbar\Omega/e$ between two superconductors. Assuming a symmetric capacitive coupling between the dot and the two superconductors, the voltage drop and, subsequently, the time-dependent phases of the superconductors are divided symmetrically, $\phi_{1,2}=\pm \Omega t$, such that
\begin{equation}
{\Delta}_{\rm eff}(t)=\frac{1}{2}\left(\Gamma_{S1}e^{i\Omega t}+\Gamma_{S2}e^{-i\Omega t}\right).
\end{equation}
The two tunneling rates $\Gamma_{S1,2}$ can, in general, differ from each other. 
In the second scheme, shown in Fig. \ref{fig1}(b),  the quantum dot is coupled to one normal and one superconducting lead with $\Gamma_S(t)=\bar \Gamma_{S} +\delta \Gamma_S \cos(\Omega t)$, yielding 
${\Delta}_{\rm eff}(t)=\Gamma_S(t)/2$. 
In the following, we formulate the results for the first scheme. 
The second scheme is covered by substituting $\Gamma_{S1}$ and $\Gamma_{S2}$ with $\delta\Gamma_{S}/2$.
\par
In order to obtain an effective description of the dot, we trace out the normal lead's degrees of freedom. This leads to the reduced density matrix of the dot $\rho_{\rm QD}$ with matrix elements $P^{\alpha}_{\beta}(t)=\langle \alpha|\rho_{\rm QD}(t)|\beta\rangle$ with respect to the basis states $|0\rangle, |\uparrow\rangle, |\downarrow\rangle, |d\rangle\equiv d^\dag_{\downarrow}d^\dag_{\uparrow}|0\rangle$ for the empty, singly-occupied with spin up or down, and doubly-occupied dot. The corresponding energies are $E_0=0$, $E_{\downarrow}=E_{\uparrow}=\epsilon$, and $E_{d}=2\epsilon+U=\delta$, respectively. 
Among the off-diagonal elements,  only the dynamics of $P_{0}^{d}$ and $P^{0}_{d}$  are coupled to the occupation probabilities. The dot's pair amplitude is described by $P^{d}_{0}(t)=\langle d_{\downarrow}(t)d_{\uparrow}(t)\rangle$. 
The time evolution of the reduced density matrix of the dot after time $0$, at which the driving is switched on, is governed by the kinetic equation,
\begin{eqnarray}\label{master}
\frac{d}{dt} P^{\alpha}_{\beta}(t)&+&i [H_{\rm QD,eff}(t), \rho_{\rm QD}(t)]^{\alpha}_{\beta}  \nonumber\\
&=&\sum_{\alpha',\beta'}  \int_{0}^{t}dt'   W_{\beta\beta'}^{\alpha\alpha'}(t,t')    P_{\beta'}^{\alpha'}(t'),
\end{eqnarray}
where $W_{\beta\beta'}^{\alpha\alpha'}(t,t')$ are transition rates due to the coupling to the normal lead. These rates can be evaluated by means of a real-time diagrammatic technique.~\cite{governale08}
\par
The evaluation of the diagrams is considerably simplified by employing the rotating-wave approximation, which is justified in the limit $|\delta|, |\Omega| \gg |{\Delta}_{\rm eff}(t)|$.
It consists of neglecting the non-resonant component, i.e., of replacing ${\Delta}_{\rm eff}(t) \rightarrow \frac{1}{2} \Gamma_{S1}e^{i\Omega t}$ or ${\Delta}_{\rm eff}(t) \rightarrow \frac{1}{2} \Gamma_{S2}e^{-i\Omega t}$ if $\delta$ and $\Omega$ have the same or opposite sign, respectively.
In the following, we write all formulas for the first case.
The second one is obtained by replacing $\Gamma_{S1}$ with $\Gamma_{S2}$ and $\Omega$ with $-\Omega$.
Within in the rotating-wave approximation the time evolution due to $H_{\rm QD,eff}$, which enters the calculations of the rates $W_{\beta\beta'}^{\alpha\alpha'}(t,t')$, takes the form 
\begin{align}
U(t,t')&=e^{-i\epsilon (t-t')}{\Big(}\sum_{\sigma}|\sigma\rangle \langle \sigma|+\alpha_{+}|0\rangle \langle0|+\alpha_{-}|d\rangle \langle d|\nonumber\\
&+\beta_{+}|0\rangle \langle d|-\beta_{-}|d\rangle \langle 0|{\Big)}\\
\alpha_{\pm}(t,t')&=\frac{1}{2} \sum_{\gamma=\pm} \left(1+\gamma\frac{\delta-\Omega}{2\epsilon_{A}}\right)  
e^{\pm i\varepsilon_{\gamma,\mp} (t-t')} \\
\beta_{\pm}(t,t')&= \sum_{\gamma=\pm}  \gamma  \frac{\Gamma_{S1}}{4\epsilon_{A}} 
e^{\pm i\Omega t}  e^{\mp i\varepsilon_{\gamma,\pm}(t-t')}
\end{align}
Here 
$\varepsilon_{\gamma\gamma'}=\gamma\epsilon_{A}+(\gamma' U +\Omega)/2$ are Andreev addition energies with $\gamma,\gamma'=\pm 1$  and $2\epsilon_{A}=\sqrt{(\delta-\Omega)^2+\Gamma_{S1}^2}$. 
\par
By explicit calculation, we find that the kinetic equations for $P_\uparrow$, $P_\downarrow$, and $P_d+P_0$ decouple from those for $P^{d}_{0} = \left(P^{0}_{d}\right)^*$ and $P_d-P_0$.
The latter are conveniently expressed in terms of an isospin ${\bf I}=({\rm Re} \, P^{d}_{0}, {\rm Im} \, P^{d}_{0}, (P_{d}-P_{0})/2 )$.
To first order in $\Gamma_N$, which describes the regime of weak tunnel coupling between dot and normal lead ($\Gamma_N\ll k_BT$), the isospin dynamics is governed by  
\begin{align}
\frac{d}{dt}{\bf I}(t)&-{\bf I}(t)\times{\bf B}^{(0)}(t) = \nonumber\\
&\Gamma_N\int^{t}_{0} dt' \left[{\bf I}(t') \times{\bf B}  -\hat{\bf R} \cdot {\bf I}(t') + {\bf A}  \right].\label{isospin}
\end{align}
Here, ${\bf B}^{(0)}(t)=(\Gamma_{S1}\cos(\Omega t),\Gamma_{S1}\sin(\Omega t),-\delta)$
is the effective magnetic field inducing a fully-coherent evolution of the isospin in the absence of the normal lead.
A finite tunnel coupling to a normal lead (r.h.s. of Eq.~(\ref{isospin})) introduces a correction to the isospin rotation about vector ${\bf B}(t,t')$, a relaxation matrix $\hat{\bf R}(t,t')$ and a vector ${\bf A}(t,t')$ associated with the generation of the isospin.
Their (real) components are determined by,
\begin{align}
&B_x+iB_y = -i  \int \frac{d\omega}{\pi}  (\beta_{+}-\beta_{-}^{\ast})  f(\omega) e^{-i\omega(t-t') },\\
&R_{xx}+iB_z=  \int \frac{d\omega}{\pi}  [1-(\alpha_{+}-\alpha_{-}^{\ast})  f(\omega)] e^{i\omega(t-t') },\\
&R_{zz} ={\rm Re} \,  \int \frac{d\omega}{\pi}  [1+(\alpha_{+}-\alpha_{-}^{\ast})  f(\omega)] e^{-i\omega(t-t') },\\
&A_x+iA_y =  \int \frac{d\omega}{2\pi}  (\beta_{+}+\beta_{-}^{\ast})  f(\omega) e^{i\omega(t-t') },\\
&A_z={\rm Re} \,\int \frac{d\omega}{2\pi} [(\alpha_{+}+\alpha_{-}^{\ast})f(\omega)-1]    e^{-i\omega(t-t') }.
\end{align}
Here $f(\omega)=1/[1+\exp(\omega/k_BT)]$ indicates the Fermi function. We note that the relaxation matrix is diagonal and  $R_{yy}=R_{xx}$
\par
In the absence of a normal lead, $\Gamma_N\equiv 0$, Eq. (\ref{isospin}) describes a fully coherent evolution of the isospin under ${\bf B}^{(0)}(t)$.
The solution in this case can be written as ${\bf I}(t)=\hat{\Theta}(t){\bf I}(0)$ with $\hat{\Theta}(t)$ being a rotation matrix with time-dependent rotation axis and angle. 
To solve Eq. (\ref{isospin}) in the presence of the normal lead we, first, define the auxiliary quantity ${\bf J}(t)=\hat{\Theta}^{-1}(t){\bf I}(t)$, which describes the isospin in the time-dependent coordinate system that follows the coherent dynamics induced by ${\bf B}^{(0)}(t)$.
The kinetic equation for ${\bf J}(t)$ is still an integro-differential equation.
However, when restricting to first-order tunneling processes, we can substitute ${\bf J}(t)$ for ${\bf J}(t')$ in the integrand.\cite{braggio06}
In fact, the master equation shows that the derivatives appearing in the general expansion ${\bf J}(t')=\sum_{n=0}^{\infty}(t-t')^{n}[d^{n}{\bf J}(t)/dt^{n}]/n!$ are at least of first order in $\Gamma_N$.
Plugging them into the integrand on the r.h.s. of the master equation for ${\bf J}(t)$, these terms correspond to at least second order in $\Gamma_N$ and are, therefore, neglected.
Another simplification we make is the replacement of the lower limit of the integral by $-\infty$ which is justified when $t$ is much larger than the characteristic decay time of the integral kernel. 
Explicitly performing the integral leads to a pure differential equation for ${\bf J}(t)$ that can be solved analytically.
This yields lengthy expressions that we do not show here.
The time evolution of ${\bf I}(t)$ is, then, obtained by going back to the original coordinate system.
\par
\section{Results}
In the following, we consider two different regimes: (1) $\Gamma_S\gg\Gamma_N$ and (2) $\Gamma_S\sim\Gamma_N$. In the first case, when the dot is strongly coupled to the superconductor(s) and weakly coupled to the normal lead, the system is mostly governed with Rabi-resonance physics. It is in this regime, that coherent manipulations of the dot charge states, detectable by coherent oscillations in the current, can be achieved. In the second regime, the Rabi resonance still plays an important role, but features related to quantum stochastic resonance become relevant since the driving and relaxation have the same order.\\

\subsection{Strong coupling to the superconductor}
Here we investigate the regime in which the coupling of the dot to the superconductors is much stronger than to the normal lead ($\Gamma_S \gg \Gamma_N$). 
First, we discuss the steady-state behavior when the system reaches a constant ${\bf J}= {\bf J}_{\infty}$ for times $t\gg 1/\Gamma_N$. The corresponding isospin, ${\bf I}_{\infty}(t)=\hat{\Theta}(t){\bf J}_{\infty}$, rotates about the $z$-axis with frequency $\Omega$, i.e., in addition to the constant $z$-component, there is an oscillatory in-plane part of the isospin. \\
In the limit of large $\Omega,\delta$ when $f(\varepsilon_{\gamma\gamma'})$ for all Andreev addition energies are approximately zero, the isospin can be written as
\begin{equation}
{\bf I}(t)=\frac{1}{2}\frac{\Gamma_{S1}(\Omega-\delta)}{(\delta-\Omega)^2+\Gamma_{S1}^2}\left(\cos\Omega t, \sin \Omega t,\frac{\delta-\Omega}{\Gamma_{S1}}\right)\label{isospin-strong}
\end{equation}
The in-plane components of the isospin indicate superconducting correlations, described by the pair amplitude
$\langle d_{\downarrow}d_{\uparrow}\rangle=P^{d}_{0}=I_x+iI_y\equiv e^{i\Omega t} \Delta_{\rm QD}$, where
$\Delta_{\rm QD}$ characterizes the (time-independent) strength of the proximity effect.
For large detuning, $|\delta|\gg \Gamma_{S1}$, the $z$-component of ${\bf B}^{(0)}$ tries to keep the isospin aligned with the $z$-direction, which corresponds to a negligibly small $\Delta_{\rm QD}$. 
However, once the driving frequency fulfills the resonance condition $|\delta - \Omega|\lesssim \Gamma_{S1}$, the in-plane component of the isospin becomes significant.
If $\delta$ and $\Omega$ have different sign then the resonance condition is replaced by $|\delta + \Omega|\lesssim \Gamma_{S2}$.
\begin{figure}[tp]
	\includegraphics[width=8cm]{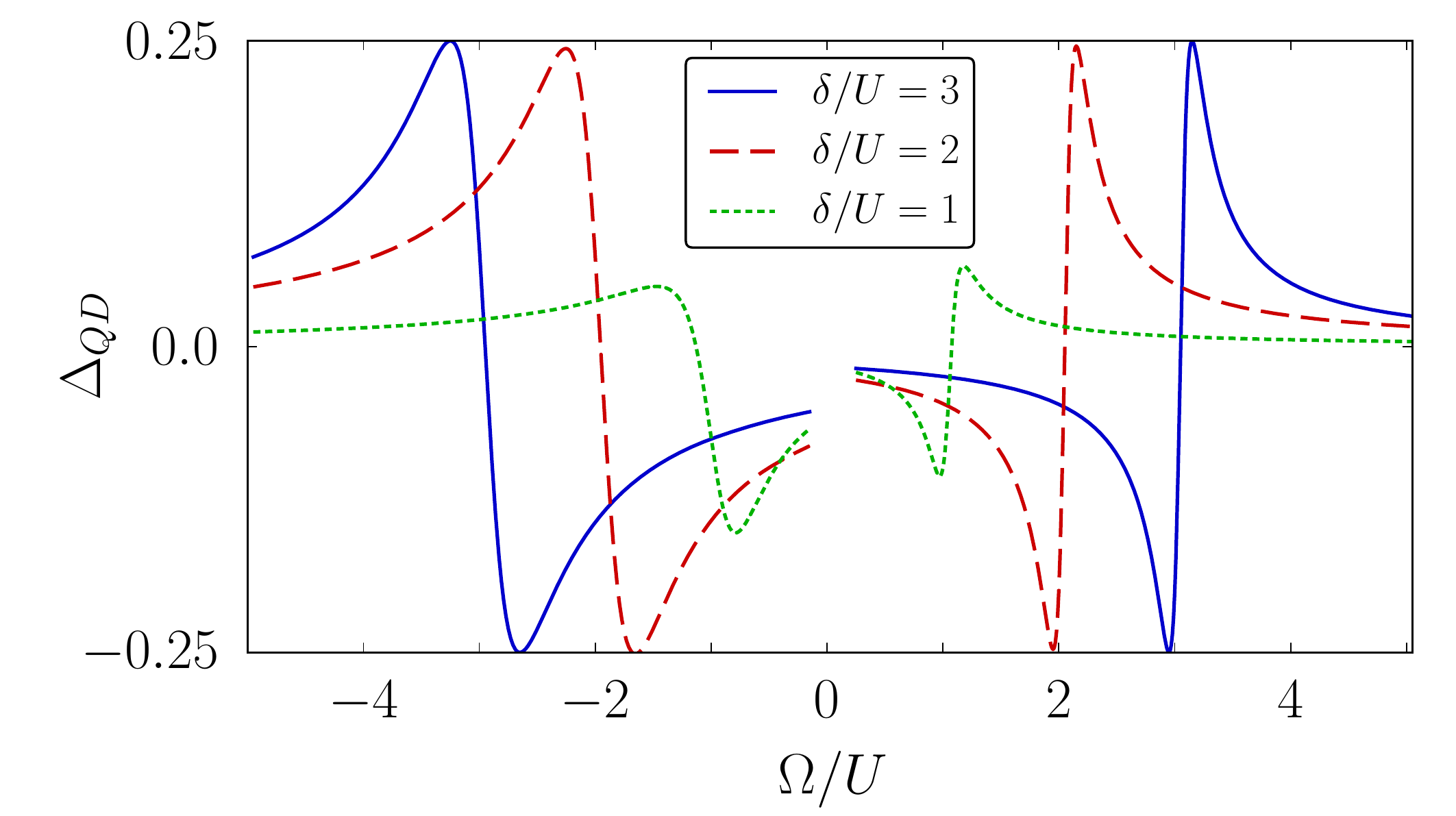}
	\caption{\label{fig2}(Color online)   Pair amplitude strength $\Delta_{\text{QD}}$ as a function of oscillation frequency $\Omega$ for different detunings $\delta$. The parameters are ${\Gamma}_{S1}=1 k_BT$,  ${\Gamma}_{S2}=3 k_BT$, and $U=10 k_BT$. Strong pairing is seen around the resonance points $\Omega\sim\pm\delta$. The apparent discontinuity at $\Omega=0$ is due to the failure of the rotating-wave approximation at small frequencies.  }
\end{figure}
Figure \ref{fig2} shows $\Delta_{\text{QD}}$ as a function of driving frequency $\Omega$ for various values of detuning $\delta$.
Close to the resonance conditions and more precisely at $\Omega-\delta=\pm \Gamma_{S1}$ and $\Omega+\delta=\mp \Gamma_{S2}$, the pair amplitude attains the maximum and minimum values of $\pm1/4$ as one can see from the isospin solution given by Eq.~(\ref{isospin-strong}).
Exactly at resonance, $\Delta_{\text{QD}}$ displays an abrupt sign change, which is reminiscent of the  
$0-\pi$ transition in the Josephson current.\cite{kouwenhoven06,governale08} 
Furthermore, we remark that for $|\delta| \lesssim U$ the pair amplitude is strongly suppressed since in this regime single occupation of the dot is favored. 
\par
The nonequilibrium proximity effect can be probed by measuring the current into a normal lead, tunnel-coupled to the dot. 
The current in lead $N$ can be computed by means of the general formula
$I_{N}(t)=\sum_{\alpha\alpha'\beta'}\int^{t}_{0}dt' W^{ N \alpha\alpha'}_{~\alpha\beta'}(t,t')P^{\alpha'}_{\beta'}(t')$, 
where the current rates $W^{N\alpha\alpha'}_{~\alpha\beta'}(t,t')$ take into account the number of charges that flow in  or out of the normal lead. In our system, the  current up to first order in $\Gamma_N$ depends only on the isospin degree of freedom and is given by
\begin{equation}
I_{ N}(t)=\frac{e}{\hbar}\Gamma_N\int_{0}^{t} dt'\left\{   R_{zz}I_z-A_z+({\bf I}\times{\bf B})_{z}   \right \}. \label{current}
\end{equation}
In the steady-state limit the current turns out to be constant and for large frequency and detuning, using the isospin given by Eq.~(\ref{isospin-strong}) we obtain the current with a Lorentzian form,
\begin{equation}
I_{ N}=\frac{e}{\hbar}\Gamma_N \frac{\Gamma_{S1}^2}{\Gamma_{S1}^2+(\delta-\Omega)^2} \,.\label{current-strong}
\end{equation}
\begin{figure}[tp]
	\includegraphics[width=8cm]{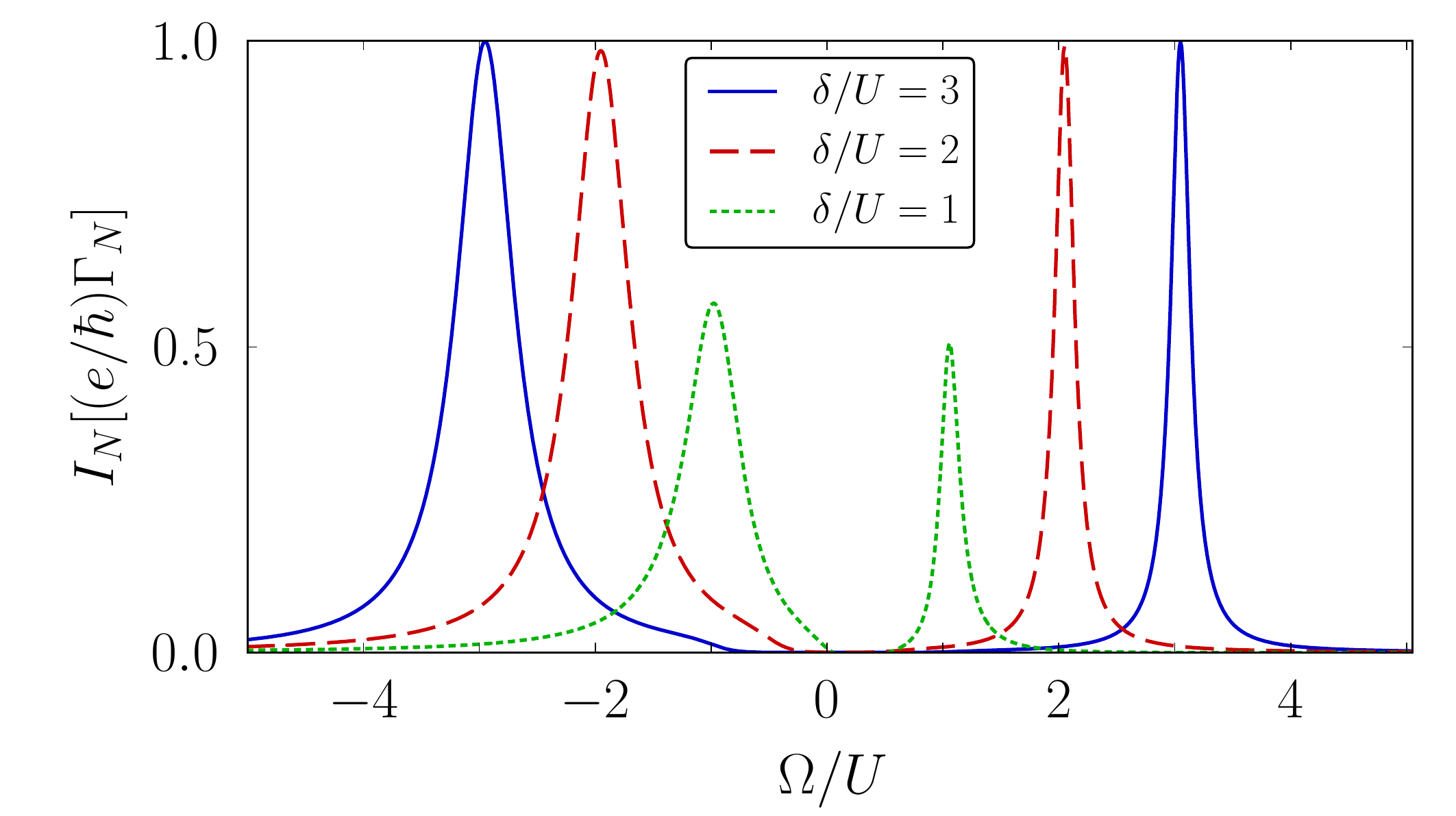}
	\caption{\label{fig3}(Color online) Normal-lead current as a function of frequency $\Omega$ for different values of $\delta$. The parameters are ${\Gamma}_{S1}=1 k_BT$,  ${\Gamma}_{S2}=3 k_BT$, and $U=10 k_BT$. The current shows resonance peaks at $\Omega=+\delta$ and $-\delta$ with widths $\Gamma_{S1}$ and $\Gamma_{S2}$, respectively. }
\end{figure}
In Fig. \ref{fig3} the current is plotted as a function of frequency for various values of $\delta>0$. 
Resonance peaks with width $\Gamma_{S1}$ and $\Gamma_{S2}$ appear at $\Omega=+\delta$ and $-\delta$, respectively. 
For negative values of $\delta$, the current flows in the opposite direction, and the position of the peaks with width 
$\Gamma_{S1}$ and $\Gamma_{S2}$ are interchanged.
\begin{figure}[tp]
	\includegraphics[width=8cm]{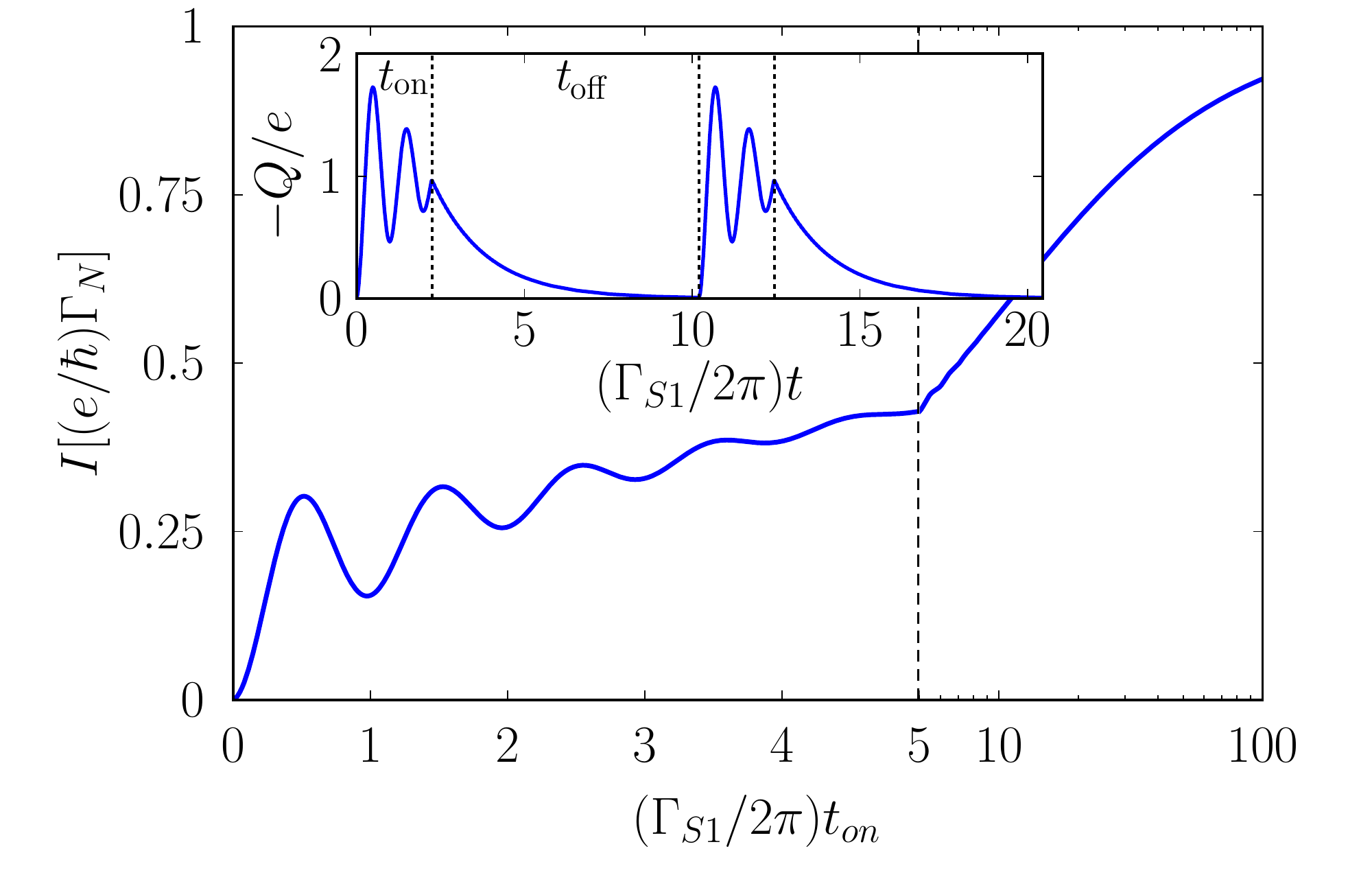}
	\caption{\label{fig4}(Color online) 
	The time-averaged current in the normal lead as a function of pulse duration $t_{\rm on}$. The inset shows the real-time variations of the dot charge when the driving is switched on and off periodically with durations $t_{\rm on}=0.45 \pi/\Gamma_N$ and $t_{\rm off}=5/\Gamma_N$, respectively. The values of the parameters used for this figure are $U=10k_BT$, $\Omega=\delta=50k_BT$, $\Gamma_{S1}=\Gamma_{S2}=2 k_BT$, and $\Gamma_N=0.2 k_BT$.		}
\end{figure}
\par
In order to discuss the transient behavior of the driven dot we assume a positive large $\delta > U$ which in equilibrium, causes the dot to be empty, ${\bf I}=(0,0,-1/2)$.
After switching on the resonant driving ($\Omega=\delta$) at $t=0$, $I_z$ decays on the time scale $1/\Gamma_N$, accompanied by oscillations with frequency $\Gamma_{S1}$.
Pulses of finite length in time allow to generate and manipulate coherent superposition of the empty and doubly-occupied states. 
For instance a pulse with duration $t=\pi/(2\Gamma_{S1})$ causes a rotation of isospin from $\hat{z}$ into the $\hat{x}$-$\hat{y}$-plane. 
In order to probe these Rabi-like oscillations via the (intrinsically stochastic) tunnel current into the normal lead, we propose to periodically switch the driving on and off with durations $t_{\rm on}$ and $t_{\rm off}$ (where $t_{\rm off}\gg 1/\Gamma_N$).
The time-averaged current into the normal lead then shows damped oscillations as function of $t_{\rm on}$ and reaches the steady state value $e\Gamma_N/\hbar$ for $t_{\rm on}\gg t_{\rm off}$, see Fig.~\ref{fig4}. The underlying real-time dynamics of the dot charge $Q=-e(1+2I_z)$ shows damped Rabi-like oscillations (pure decay) when the driving is switched on (off) as illustrated in the inset of Fig.~\ref{fig4}. 

\subsection{Weak coupling to the superconductor}
\begin{figure}[tp]
	\includegraphics[width=8cm]{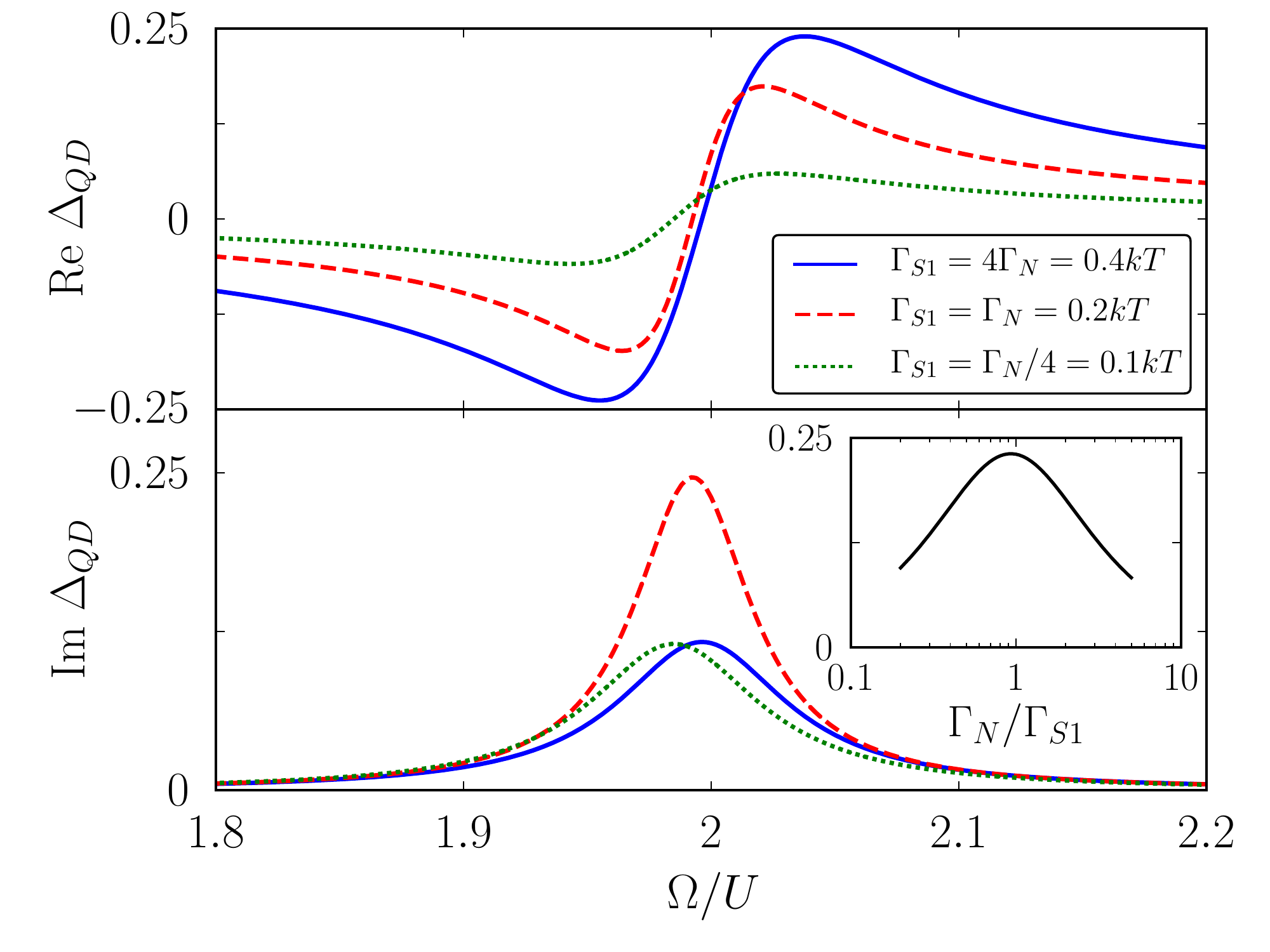}
	\caption{\label{fig5}(Color online) 
	Real and imaginary components of the pairing strength $\Delta_{QD}$ as a function of frequency $\Omega$
	for various values of tunnel couplings $\Gamma_N$ and $\Gamma_{S1}$. 
	The inset of the lower panel shows the dependence of the imaginary part
	of $\Delta_{QD}$ on the relative value of the tunnel couplings $\Gamma_N/\Gamma_{S1}$ assuming $\Omega=\delta$ and 
	$\Gamma_{S1}=0.1 k_BT$. The values of parameters used in this figure are $U=10 k_BT$ and $\delta=2U$. 
	Here the pairing strength around $\Omega\sim\delta$ is shown and for negative values $\Omega\sim-\delta$ the plots 
	can be obtained by mirror reflections with respect to $0$ and with widths determined by $\Gamma_{S2}$. 
	}
\end{figure}
We now turn to the regime of weak coupling to both superconducting and normal leads ($\Gamma_S\sim\Gamma_N \ll k_BT$). 
For a systematic perturbative treatment of the isospin dynamics, we count both $\Gamma_S$ and $\Gamma_N$ but also the detuning from the resonance condition $\delta-\Omega$ as one order in the perturbation expansion.
As a consequence, we drop all higher-order terms of $\Gamma_S$ and $\delta-\Omega$ in the master equation.
This can be easily achieved by the replacement $\varepsilon_{\gamma\gamma'}(\epsilon_A\to 0)=(\gamma'U+\Omega)/2$ for the Andreev addition energies.
As a result the form of the kinetic equation (\ref{isospin}) for the isospin remains unchanged but ${\bf B}$, $\hat{\bf R}$, and ${\bf A}$ are modified according to the replacement of Andreev addition energies.
In addition the effective magnetic field ${\bf B}^{(0)}$ (zeroth order with respect to $\Gamma_N$) should now be considered as sum of a zeroth-order, $(0,0,-\delta)$, and a first-order part, $(\Gamma_{S1}\cos(\Omega t),\Gamma_{S1}\sin(\Omega t),0)$, with respect to $\Gamma_S$. Therefore in this case the matrix $\hat{\Theta}$ used in the definition of auxiliary quantity ${\bf J}=\hat{\Theta}^{-1} {\bf I}$ corresponds to the coherent evolution under $(0,0,-\delta)$. 
\par
Despite these modifications, the method to obtain the isospin dynamics is the same as in the strong coupling case.
For the steady-state limit, we find
\begin{align}
{\bf I}(t)&=\frac{{\mathcal A}/{\mathcal R}}{ {\mathcal R}^2+\Gamma_{S1}^2 +(\delta-\Omega-{\mathcal B})^2 } \nonumber\\
&\times \left(
\begin{matrix}
\Gamma_{S1}[(\delta-\Omega-{\mathcal B})\cos\Omega t+{\mathcal R}\sin\Omega t]
\cr
\Gamma_{S1}[(\delta-\Omega-{\mathcal B})\sin\Omega t-{\mathcal R}\cos\Omega t]
\cr
{\mathcal R}^2+(\delta-\Omega-{\mathcal B})^2 \end{matrix}
\right ),\label{isospin-weak}
\end{align}
with
\begin{align}
&{\mathcal R}=\Gamma_N\left[f\left(\frac{\Omega -U}{2}\right)+f\left(\frac{-\Omega -U}{2}\right)\right]
\\
&{\mathcal A}=\frac{\Gamma_N}{2}\left[f\left(\frac{\Omega -U}{2}\right)-f\left(\frac{-\Omega -U}{2}\right)\right]
\\
&{\mathcal B}=\frac{\Gamma_N}{\pi}{\rm Re}\, \left[\psi\left(\frac{1}{2}+i\frac{\Omega+U}{4\pi k_BT}\right)-\psi\left(\frac{1}{2}+i\frac{\Omega-U}{4\pi k_BT}\right)\right]
\end{align}
in which $\psi$ is the digamma function. 
Here, we follow the same convention for the rotating-wave approximation as before: only the results for the case of $\delta$ and $\Omega$ having the same sign are presented.
The results for $\delta$ and $\Omega$ having opposite signs can be obtained by the replacement $\Omega\to -\Omega$ and $\Gamma_{S1}\to \Gamma_{S2}$.
From the isospin form given above, we 
find an oscillatory pair amplitude with frequency $\Omega$ and a complex-valued strength
\begin{equation}
\Delta_{\rm QD}=\Gamma_{S1}\frac{\mathcal A} {\mathcal R} \frac{\delta-\Omega-{\mathcal B}-i{\mathcal R}}{ {\mathcal R}^2+\Gamma_{S1}^2 +(\delta-\Omega-{\mathcal B})^2 }. \label{Delta-weak}\end{equation}
Similar to the strong-coupling limit, the pairing strength has a resonant form as function of $\Omega$.
There are, however, differences. First, $\Delta_{\rm QD}$ now depends on both couplings $\Gamma_{S1}$ and $\Gamma_N$ and, second the resonance point is slightly displaced by an effective exchange field ${\mathcal B}$.  
As one can see from Fig. \ref{fig5} the real part of the pairing strength behaves in a similar way to $\Delta_{\rm QD}$ in the strong-coupling limit and passes through a maximum/minimum at $\delta-\Omega-{\mathcal B}=\pm\sqrt{ {\mathcal R}^2+\Gamma_{S1}^2 }$ accompanied with a sign change at $\delta-\Omega-{\mathcal B}=0$. On the other hand, the imaginary part, which was absent in the strong-coupling regime, shows a single resonance peak at $\delta-\Omega-{\mathcal B}=0$.
Interestingly, although the maximum value of the real part monotonically decreases by increasing $\Gamma_N$, the peak value of imaginary part has a maximum for $\Gamma_N=\Gamma_{S1}$.
This is clearly seen in the inset of Fig. \ref{fig5} and can be understood as a manifestation of the so-called quantum stochastic resonance,\cite{gammaitoni,wellens} a phenomenon which appears when the response of the system to a driving signal is increased by a certain level of dissipation\cite{viola} or noise.\cite{nori09}
We note that away from the resonance point, the above-mentioned signature of the quantum stochastic resonance disappears: the absolute value of the pairing, which is physically more relevant, does not need an optimal value of $\Gamma_N$ to become large. As one can easily check from Eq.~(\ref{Delta-weak}),  $|\Delta_{\rm QD}|$ passes through a maximum $|{\mathcal A}|/2{\mathcal R}$ when $(\delta-\Omega-{\mathcal B})^2+ {\mathcal R}^2=\Gamma_{S1}^2 $. This is not achievable for large dissipation $\Gamma_N\gg \Gamma_{S1}$, but always possible for $\Gamma_{S1}>\Gamma_N$, with an optimal pairing amplitude of $1/4$, in accordance with the result for the strong-coupling regime.
\par
To obtain the current we use (\ref{current}) and introduce the same modifications as the for kinetic equation.  
We plug in the isospin solution (\ref{isospin-weak}) and get the steady-state current,
\begin{equation}
I_N=\frac{e}{\hbar} \frac{2\Gamma_{S1}^2 {\mathcal A}}{ {\mathcal R}^2+\Gamma_{S1}^2 +(\delta-\Omega-{\mathcal B})^2} \, ,
\label{current-weak}
\end{equation}
which, again, has a Lorentzian form with displaced resonance $\delta-\Omega={\mathcal B}$ and width $\sqrt{{\mathcal R}^2+\Gamma_{S1}^2}$. 
For $\Gamma_{S1}\gg\Gamma_N$, the current reaches the value Eq~(\ref{current-strong}) of the strong-coupling regime. 
\par
We end up this subsection by commenting on the transient behavior in this regime. Since we are assuming weak coupling to the superconductor $\Gamma_S\sim\Gamma_N$, the relaxation mechanism dominates over the coherent evolution produced by the driving and no coherent oscillations can be seen.
This means that a coherent-manipulation scheme of the charge states requires a stronger coupling to the superconductors than to the normal lead. 
\par
Finally, we comment on possible experimental verification of the predicted effects. Considering typical experimental values $k_BT \sim 1\ \mu$eV, $|\Delta|\sim0.15$ meV (for Al/Ti)~\cite{kouwenhoven06}, $\delta$ and $\Omega$ must be on the order of $10\ \mu$eV to fulfill the requirement $\Gamma_N\ll k_BT, \Gamma_S \ll |\delta|,|\Omega| \ll |\Delta|$.
Therefore, for the setup where magnetic flux is used to produce a time-depedent tunnel coupling, a magnetic field variation rate $dB/dt \sim 10^5$ T/s is necessary for a superconducting loop with area $100\  \mu$m$^2$. 

\section{Conclusions}
We have proposed the possibility to induce superconducting correlations in a single-level quantum dot via a time-dependent tunnel coupling to a superconducting lead.
A finite pair amplitude can be generated if a resonance condition for the driving frequency in relation to the energy difference between empty and doubly-occupied dot is fulfilled. 
The existence of this nonequilibrium proximity effect can be probed by measuring a finite steady-state current into a weakly-coupled normal lead.
Furthermore, Rabi-like oscillations between empty and  doubly-occupied dot states may be probed  in the tunnel current by applying periodically pulsed oscillations of the tunnel coupling to the superconductor with tunable pulse duration.

\acknowledgments
We acknowledge financial support from EU under grant No. 238345 (GEOMDISS) and the DFG via grant KO1987/5.

\end{document}